\documentclass[10pt]{article}

\pdfoutput=1

\newcommand{\mode}{arxiv}
\usepackage{ifthen}

\newcommand{\plos}{\equal{\mode}{plos}}

\usepackage{amsmath}
\usepackage{amssymb}

\usepackage{graphicx}

\usepackage{cite}

\usepackage{color} 

\usepackage{setspace}
\ifthenelse{\plos}
{\doublespacing}
{\singlespacing}

\usepackage[labelfont=bf,labelsep=period,justification=raggedright]{caption}

\makeatletter
\renewcommand{\@biblabel}[1]{\quad#1.}
\makeatother

\date{}

\usepackage{lineno}
\usepackage{multicol}

\newcommand{\nspecies}{170}
\newcommand{\ncarnivores}{9}
\newcommand{\nherbivores}{25}
\newcommand{\nplants}{136}
\newcommand{\nlinks}{667}
\newcommand{\linkdensity}{0.0231}
\newcommand{\nherbivory}{550}
\newcommand{\npredatory}{117}
\newcommand{\mlmindiff}{106}
\newcommand{\mlmaxdiff}{470}
\newcommand{\chimean}{3.2}
\newcommand{\chileft}{1.6}
\newcommand{\chiright}{5.3}
\newcommand{\kmean}{14.9}
\newcommand{\kleft}{12}
\newcommand{\kright}{18}
\newcommand{\kpriormeanu}{43.8}
\newcommand{\kpriormeand}{5.4}
\newcommand{\alphamean}{0.046}
\newcommand{\alphaleft}{0.029}
\newcommand{\alpharight}{0.068}
\newcommand{\betamean}{0.89}
\newcommand{\betaleft}{0.50}
\newcommand{\betaright}{1.43}
\newcommand{\kconsensus}{16}
\newcommand{\nplantsmean}{22.7}
\newcommand{\nherbivoresmean}{3.6}
\newcommand{\ncarnivoresmean}{3.0}
\newcommand{\gCROCRO}{1}
\newcommand{\gCANAUR}{2}
\newcommand{\gCARCAR}{3}
\newcommand{\gCONTAU}{4}
\newcommand{\gREDRED}{5}
\newcommand{\gSYNCAF}{6}
\newcommand{\gHETBRU}{7}
\newcommand{\gAGAPLA}{8}
\newcommand{\gGIRCAM}{9}
\newcommand{\gLOXAFR}{10}
\newcommand{\gTHETRI}{11}
\newcommand{\gERACIL}{12}
\newcommand{\gMICKUN}{13}
\newcommand{\gANDSCH}{14}
\newcommand{\gACASEN}{15}
\newcommand{\gCRODIC}{16}

\newcommand{\like}{\mathcal{L}}
\newcommand{\bell}{\mathcal{B}}
\newcommand{\stirling}{\mathcal{S}}

\newcommand{\pr}{\mathrm{Pr}}

\usepackage{ulem}
\definecolor{blue}{rgb}{0.324,0.609,0.708}

\definecolor{purple}{rgb}{0.459,0.109,0.538}

\definecolor{orange}{rgb}{1,0.5,0}

\definecolor{red}{rgb}{0.6,0.1,0.3}

\definecolor{green}{rgb}{0.513,0.73,0.442}

\begin{document}


\begin{flushleft}
{\Large
\textbf{Spatial Guilds in the Serengeti Food Web Revealed by a Bayesian Group Model}}
\\
Edward B. Baskerville$^{1,\ast}$,
Andy P. Dobson$^{2}$, 
Trevor Bedford$^{1,3}$,
Stefano Allesina$^{4}$,
Mercedes Pascual$^{1,3}$
\\
\bf{1} Department of Ecology and Evolutionary Biology, University of
Michigan, Ann Arbor, Michigan, United States of America
\\
\bf{2} Department of Ecology and Evolutionary Biology, Princeton University,
Princeton, New Jersey, United States of America
\\
\bf{3} Howard Hughes Medical Institute, University of Michigan, Ann Arbor,
Michigan, United States of America
\\
\bf{4} Department of Ecology and Evolution, Computation Institute, The
University of Chicago, Chicago, Illinois, United States of America
\\
$\ast$ E-mail: ebaskerv@umich.edu
\end{flushleft}

\section*{Abstract}

Food webs, networks of feeding relationships among organisms, provide fundamental insights into mechanisms that determine ecosystem stability and persistence. Despite long-standing interest in the compartmental structure of food webs, past network analyses of food webs have been constrained by a standard definition of compartments, or modules, that requires many links within compartments and few links between them. Empirical analyses have been further limited by low-resolution data for primary producers. In this paper, we present a Bayesian computational method for identifying group structure in food webs using a flexible definition of a group that can describe both functional roles and standard compartments. The Serengeti ecosystem provides an opportunity to examine structure in a newly compiled food web that includes species-level resolution among plants, allowing us to address whether groups in the food web correspond to tightly-connected compartments or functional groups, and whether network structure reflects spatial or trophic organization, or a combination of the two. We have compiled the major mammalian and plant components of the Serengeti food web from published literature, and we infer its group structure using our method. We find that network structure corresponds to spatially distinct plant groups coupled at higher trophic levels by groups of herbivores, which are in turn coupled by carnivore groups. Thus the group structure of the Serengeti web represents a mixture of trophic guild structure and spatial patterns, in contrast to the standard compartments typically identified in ecological networks. From data consisting only of nodes and links, the group structure that emerges supports recent ideas on spatial coupling and energy channels in ecosystems that have been proposed as important for persistence. Our Bayesian approach provides a powerful, flexible framework for the study of network structure; we believe it will prove instrumental in a variety of biological contexts.

\ifthenelse{\plos}
{\section*{Author Summary}
The relationships among organisms in an ecosystem can be described by a food web, a network representing who eats whom. Food web organization has important consequences for how populations change over time, how one species extinction can cause others, and generally how robustly ecosystems respond to disturbances. We present a computational method to analyze how species are organized into groups based on their interactions. We apply this method to the plant and mammal food web from the Serengeti grassland ecosystem in Tanzania, a pristine ecosystem increasingly threatened by human impacts. This web is unusually detailed, with plants identified down to individual species and corresponding habitats. Our analysis reveals that the Serengeti web is divided into distinct groups of plants in several habitats, which are fed on by distinct groups of herbivores, which are in turn fed on by a few groups of carnivores. In other words, the structure reflects a combination of functional groups and spatial habitats, and differs significantly from the compartmental structure previously considered in ecological studies. This pattern may partly explain how the ecosystem remains in balance. Additionally, our method can be easily applied to other kinds of networks and modified to find other network patterns.
}
{}

\ifthenelse{\plos}
{}
{\begin{multicols}{2}}

\section*{Introduction}

Food webs, networks of feeding relationships in ecosystems, connect the biotic
interactions among organisms with energy flows, thus linking together population
dynamics, ecosystem function, and network topology. Ecologists have been using
this powerful conceptual tool for more than a century \cite{cohen1995, camerano1880, elton1927}. One element of structure of particular relevance to large food webs is the
subdivision of species into compartments or groups, a feature that has been
proposed to contribute to food web stability by constraining the propagation of
disturbances through a network \cite{may1973}. Although a large literature has
considered the presence and dynamic significance of compartments in food webs, on the
whole, evidence for the importance of compartments has been inconclusive
\cite{pimm1980, krause2003, guimera2010, allesina2009}. In this literature,
compartments are alternately referred to as modules, clusters, or
``communities'' \cite{girvan2002}, and are defined by high link
density within groups and low link density between them.

Recent work with a probabilistic model considers a more flexible notion of
groups, allowing link density to be high or low within any group or between any
pair of groups \cite{allesina2009}. Groups can thus represent compartments in
the previous sense, but can also represent trophic guilds or roles
\cite{burns1989,luczkovich2003}, sets of species that feed on, and are fed on,
by similar sets of species. By fitting models of this type to data, the dominant
topological pattern in the network can be found, which may include (spatial)
compartments, or trophic guilds, or some combination of the two. The initial
application of this model to empirical food webs from different ecosystems has
revealed a predominance of trophic guilds rather than compartments
\cite{allesina2009}.

Two major challenges limit the application of this model in resolving the group
structure of large food webs and interpreting its biological basis. First, most
food webs have poor resolution of primary producers; plants in terrestrial
systems and phytoplankton in aquatic ones are typically represented by a few
nodes that are highly aggregated taxonomically. This is unfortunate, because
their position at the base of a food web means that primary producers are
essential for understanding how the web is spatially organized, and how this
spatial organization percolates up through the trophic relationships of other
species. Fully resolved primary producers are also essential in examining how
the structure of other trophic levels cuts across their spatial distribution,
and, in so doing, couples different habitats.

Second, some technical problems have hindered the use of probabilistic models in
analyzing group structure. Early food web models served as null models for food
web structure and were tested by generating model webs and comparing summary
statistics against data from real webs \cite{cohen1990, williams2000}. More
recently, a more rigorous approach for measuring the goodness of fit of a model
has been provided by maximum likelihood and model selection \cite{allesina2008,
allesina2009}; two problems still remain within this framework. One is
technical: standard model-selection criteria are not applicable to ``discrete
parameters'' such as group membership. The second problem is more fundamental:
there are many almost equally good arrangements, and it is desirable to extract
information not just from a single best arrangement, but also from the rest of
the ensemble.

The Bayesian approach is gaining popularity in ecological modeling due to the
philosophical and conceptual appeal of explicitly considering uncertainty in
parameter estimation as well as its methodological flexibility
\cite{mccarthy2007}. This approach is especially well-suited for handling
uncertainty in complex food web models, and allows us to overcome the
limitations of the previous implementation of the group model. In network
inference, there are only a few examples of complete Bayesian models
\cite{hoff2002, park2010} and a few examples of MCMC for maximum-likelihood
inference \cite{clauset2008, williams2010}, but Bayesian inference in
phylogenetics has been long established \cite{yang1997, mau1999}, and provides a
clear methodological analogue.

In this paper, we address the group structure of a newly assembled food web for
the mammals and plants of the Serengeti grassland ecosystem of Tanzania; we use
a new and novel approach to the identification of groups based on Bayesian
inference. We specifically ask whether the structure that emerges reflects the
underlying spatial dimension, as delineated by the different plant communities
that characterize different sub-habitats within the ecosystem, or whether it is
determined by trophic dimensions in the form of species guilds that share
functional roles.

The Serengeti food web is emerging as the most highly resolved terrestrial web
to date \cite{dobson2009}. The Serengeti has been studied as an integrated
ecosystem for almost five decades \cite{sinclair1979, sinclair1995,
sinclair2008}, and because of widespread popular familiarity with the
consumer-resource dynamics of lions, hyenas, wildebeest, zebra and grasses, it
provides a strong intuitive test for probabilistic food web models. Most
importantly, at the primary producer level, the Serengeti food web includes a
number of distinct grass and woodland plant communities on different soils and
across a rainfall gradient \cite{mcnaughton1983}. The sequential and
well-documented changes in the underlying plant diversity allows us to examine
the extent to which grassland communities define network topology at higher
trophic levels. The high resolution of the species that comprise the mammal and
plant communities of the ecosystem allow us to address the role of space in the
group structure of the food web.

From a complex network of only nodes and links that represent species and
their interactions, the groups that emerge from an otherwise blind
classification of species make remarkable biological sense. We find that the
group structure identified in the Serengeti food web represents a mixture of
trophic guild structure and spatial patterns, an arrangement that differs
significantly from the compartmental structure typically identified in
ecological networks.

\section*{Results/Discussion}

\subsection*{Bayesian Inference and Model Selection for Food Webs}

\subsubsection*{Probabilistic Models for Food Webs}

In this paper, we use probabilistic modeling as a tool for formalizing
hypotheses about food web structure. We treat a food web, an observed network of
who eats whom in an ecosystem, as data. We start with the basic question:
assuming a probabilistic model of food web structure, what is the probability of
observing this particular real-world food web? This probability is referred to
as the \textit{likelihood} of observing the data, given model parameters. In a
maximum-likelihood framework, the mechanical part of the inference process is to
find the set of model parameters that makes the likelihood as great as possible,
with the interpretation that this represents the best point estimate of the
underlying process.

We begin with the group model of Allesina and Pascual \cite{allesina2009}, which
was originally treated in a maximum-likelihood framework. Conceptually, this
model encodes the simple hypothesis that species can be divided into groups, and
species in the same group have statistically similar behavior: they tend to
consume species in certain groups and tend to be consumed by species in certain
groups. Specifically, the probability that a species belonging to group $i$ is
eaten by a species belonging to group $j$ is given by $p_{ij}$, and conversely,
the probability of a link being absent is $(1 - p_{ij})$. If there are $K$
groups, then a matrix $\mathbf{P}$ of $K^2$ link probabilities is required to
completely describe the relationships among all groups. The likelihood for the
whole network is the product over all pairs of species of the probability of a
link being present (if present) or absent (if absent). In the statistical
literature, this model structure is known as a stochastic block model
\cite{wang1987}. The assignment of species to groups is also an unobserved
parameter in this model, which adds a layer of difficulty to parameter
estimation. For example, in a network of $100$ species, there are approximately
$5 \times 10^{115}$ different ways to partition the network into groups (see
Methods). That is, if you had a computer that could process $10^{80}$ partitions
(as many partitions as there are atoms in the universe) every femtosecond
($10^{-15}$ s), it would take $1.5 \times 10^{13}$ years to process them all. (By
comparison, the universe is only $1.4 \times 10^{10}$ years old.)

The group model allows for a more flexible definition of groups than standard
approaches to network clustering, which find groups that have large numbers of
internal connections and relatively few connections between groups
\cite{girvan2002}. Because each $p_{ij}$ parameter may take any value between
$0$ and $1$, good model fits may result from other relationships, such as high
link density between groups and low link density within groups, and may
accommodate different relationships in different parts of the network. In
general, the best-fitting partitions will try to maximize or minimize the number
of links within specific groups and between specific pairs of groups. 

\subsubsection*{Bayesian Inference and Priors for the Group Model}

In a Bayesian framework, rules of probability are taken to govern both the data
and model parameters. Rather than finding the set of parameter values that
maximize the likelihood, the goal becomes to estimate a probability distribution
over parameters based on observed data. In this way, we can directly quantify
the uncertainty in our parameters in terms of probabilities. This permits
questions such as: what is the probability that a parameter lies in a particular
range? The name ``Bayesian'' comes from Bayes' rule, which tells us how to use
conditional probability statements to infer a \textit{posterior} distribution,
in this case, the probability distribution over parameter values conditional on
having observed the data, $\pr(\theta|D)$. If we are dealing
entirely with discrete probability distributions, Bayes' rule takes its most
intuitive form:
\begin{align}
\pr(\theta | D) &= \frac{\pr(\theta) \pr(D | \theta)}{\pr(D)}.
\end{align}
The numerator of the right-hand-side is the probability of producing the data
from the given parameters: the \textit{prior} probability of those parameters,
$\pr(\theta)$, times the probability of producing the data given those
parameters, $\pr(D | \theta)$, the likelihood. The denominator is the
\textit{marginal} probability of observing the data unconditional on the
particular parameter values at play, which is simply the sum of the
probabilities of all the different ways of producing the data using all possible
parameter values, $\pr(D) = \sum_\theta \pr(\theta) \pr(D |
\theta)$. In other words, in order to calculate the posterior probability of
parameters $\theta$, we add up all the different ways of producing the data
weighted by their probability, and then calculate what fraction of that
probability came from parameters $\theta$. From here, we will write these
quantities in more general notation, suitable for a mix of discrete and
continuous probability distributions:
\begin{align}
  \label{bayesrulesingle}
  f(\theta | D) &= \frac{f(\theta) f(D | \theta)}
  {\int_\theta f(\theta) f(D | \theta) \, d\theta}
\end{align}
where the integral sign represents a multiple integral over discrete and
continuous parameters.

In the Bayesian framework, the model includes not only the formulation of the
likelihood but also a prior distribution over parameters. With the group model,
this means defining a prior distribution over both link probabilities and
arrangements into groups (``partitions''). In general, priors may incorporate
informed knowledge about the system, but in this case we simply use them to
encode different variants of the same basic model. We use two distributions for
partitions and two distributions for link probabilities, which are combined to
form four different model variants.

The two alternative distributions for elements $p_{ij}$ of the link probability
matrix $\mathbf{P}$ are (1) a uniform distribution between 0 and 1, and (2) a
beta distribution with shape parameters $\alpha$ and $\beta$, which are in turn
governed by exponential distributions with mean $1$. With $\alpha$ and $\beta$
fixed at their means, alternative (2) reduces to a uniform distribution; at
other values, the distribution may take a uniform, convex, concave, or skewed
shape. Alternative (2) is thus structured hierarchically, with
exponential \textit{hyperpriors} for $\alpha$ and $\beta$ governing the beta
prior for elements of $\mathbf{P}$.

For partitions, we consider (1) a uniform distribution and (2) a distribution
generated by the Dirichlet process, sometimes referred to as the ``Chinese
restaurant process'' \cite{ferguson1973}. Alternative (2) is controlled by an
aggregation parameter $\chi$ that is in turn drawn from an exponential distribution with
mean $1$. The uniform distribution assigns equal prior probability to each
possible partition, irrespective of the number of groups. Because there are far more ways to
partition the network at an intermediate, but relatively high, number of groups,
the uniform prior implicitly biases the model toward that number. For example,
in the Serengeti food web, there are $\nspecies$ nodes, yielding an \textit{a
priori} expectation of $\kpriormeanu$ groups. In contrast, the hierarchically
structured Dirichlet process prior provides flexibility via the aggregation parameter
$\chi$. When $\chi$ is large, partitions tend to have many small groups; when
$\chi$ is small, partitions tend to have fewer groups, with a skewed group size
distribution.

We also consider two simple models without groups as null comparisons: (1) a directed random graph model (i.e., one group) with a uniform prior on a single link probability parameter $p$, and (2) a fully parameterized model, with each species in its own group (and a $\nspecies \times \nspecies$ link probability parameter matrix $\mathbf{P}$, also with a uniform link probability prior.

For a fuller discussion of models and prior distributions, in particular the properties of
the distribution generated by the Dirichlet process, see Methods.

\subsubsection*{Bayesian Model Selection via Marginal Likelihood}

The Bayesian framework provides a natural way to make probabilistic
inferences based on a particular model. However, we also want to be
able to choose between different models by quantifying their relative
goodness of fit. One approach to Bayesian model selection can be
framed directly in terms of Bayes' rule, mirroring the process for
estimating the posterior distribution over parameters for a single
model.

Consider two models, $M_1$ and $M_2$, to which we assign prior weight
$\pr(M_1)$ and $\pr(M_2)$. After the data has been
observed, we can calculate the posterior probability of the models
using Bayes' rule:
\begin{align}
\pr(M_1 | D) &= \frac{\pr(M_1) \pr(D | M_1)}{\pr(D)},
\end{align}
\begin{align}
\pr(M_2 | D) &= \frac{\pr(M_2) \pr(D | M_2)}{\pr(D)},
\end{align}
where the denominator is equal to the probability of observing the data
unconditional of the particular model at play, $\pr(D) =
\pr(M_1) P(D | M_1) + \pr(M_2) P(D | M_2)$. The
probabilities $\pr(D | M_1) = \int_{\theta_1} f(\theta_1) f(D |
\theta_1) \, d\theta_1$ and $\pr(D | M_2) = \int_{\theta_2}
f(\theta_2) f(D | \theta_2) \, d\theta_2$ are the marginal likelihoods
of the two models, corresponding to the denominator in Equation
\ref{bayesrulesingle}. If we give the two models equal prior weight,
then the relative posterior weight of the two models is simply given
by the marginal likelihoods. This reasoning extends naturally to any
number of models.

The ratio of the marginal likelihoods is often called the Bayes factor
\cite{jeffreys1935,jeffreys1961,kass1995}, and is equal to the
posterior odds ratio of the two models, assuming equal prior weight:
\begin{align}
B_{12} &= \frac{\pr(D | M_1)}{\pr(D | M_2)}
\end{align}
The Bayes factor provides a convenient way to compare models: if
$B_{12} = 10$, then we consider support for model $M_1$ to be ten
times stronger than model $M_2$. In AIC-based model selection, the Bayes factor
is analogous to a ratio of Akaike weights \cite{burnham2002}.

\subsubsection*{Consensus Partitions}

The output of an MCMC simulation includes a long sequence of network partitions
representing draws from the posterior distribution over partitions. As these
partitions are potentially all distinct from each other, but represent similar
tendencies of species to be grouped together, it is useful to try to summarize
the information contained in all the samples in a more compact form. One
approach is to construct a pairwise group-membership matrix for species in the
food web, with entries equal to the posterior probability that two species are
in the same group. A visual representation of this matrix can illuminate the
group structure, and a consensus partition can then be constructed from this
matrix using a simple clustering algorithm. (For more details, see Methods.)

\subsection*{Bayesian analysis of the Serengeti food web}

\subsubsection*{The Serengeti Data Set}

We compiled the food web
from published accounts of feeding links in the literature \cite{sinclair2003,
casebeer1970, cooper1999, hansen1985, murray1993, talbot1962, talbot1963,
schaller1972, caro1994, kruuk1972, mcnaughton1983, mcnaughton1985, vesey1960,
lamprecht1978}.

The compiled Serengeti food web
(Tables~\ref{tab:species_list}~and~\ref{tab:edge_list} and
Figure~\ref{fig:network_ungrouped}) consists of $L = \nlinks$ feeding links
among $S = \nspecies$ species ($\nplants$ plants, $\nherbivores$ herbivores, and
$\ncarnivores$ carnivores). $\nherbivory$ of the links are herbivorous, and
$\npredatory$ are predatory. The fraction of all possible links or connectance
($C = L/S^2$), ignoring all biological constraints, is equal to $\linkdensity$.

\subsubsection*{Performance of Model Variants}

We find unequivocal support for the use of group-based models in describing the Serengeti food web. Models with group structure have vastly greater marginal likelihoods than simple null models that ignore group structure (Table~\ref{tab:marginal_likelihood}). Furthermore, the use of both the Dirichlet process prior for network partitions and the beta
prior for link probability parameters vastly improved the fit of the basic
model. The best model variant as measured by marginal likelihood included both
the beta prior on link probabilities and the Dirichlet process prior on
partitions. The next best variant
included a uniform partition prior and beta link probability prior, followed by
the variant with Dirichlet process prior and uniform link probability prior. The
strongest variant surpassed its closest competitor by $\mlmindiff$ log-orders
in likelihood and surpassed the model with two uniform priors by $\mlmaxdiff$
log-orders in likelihood, providing unequivocal support for including both flexible priors in the model specification. Accordingly, in the remaining analysis we consider only the best model variant.

\subsubsection*{Identification of Model Parameters}

The posterior mean number of groups $K$ is $\kmean$ (95\% credible interval
$\kleft, \kright$), and the mean value of the Dirichlet process parameter $\chi$
is $\chimean$ ($\chileft, \chiright$) (Figure \ref{fig:chi_k}). The prior
expectation of $\chi$ was 1.0 and the prior expectation of $K$ was
$\kpriormeand$. The finding of posterior values substantially greater than prior
values strongly supports the presence of detailed group structure in the
Serengeti food web.

Mean values for beta distribution parameters are $\alpha = \alphamean$
($\alphaleft, \alpharight$) and $\beta = \betamean$ ($\betaleft, \betaright$) (Figure \ref{fig:alpha_beta}).
The corresponding beta prior has support concentrated near 0, since most species
do not feed on most other species (Figure \ref{fig:pij}).

\subsubsection*{Groups Identified in the Serengeti Food Web}

The structure of the Serengeti food web is best represented by groups containing
trophically similar species, subdivided by specialization on different
components of prey species, with plant species corresponding to spatially
distinct habitats. The best consensus partition, with $\kconsensus$ groups, is
shown in Table~\ref{tab:groups}. There
are 6 groups of plant species (groups 11--16), 7 groups of herbivore species
(groups 4--10) and 3 groups of carnivores species (groups 1--3). On average,
plant groups contain more species than herbivore and carnivore groups
($\nplantsmean$, $\nherbivoresmean$ and $\ncarnivoresmean$, respectively). As
evident in the pairwise group-membership matrix
(Figure~\ref{fig:similarity_matrix}), the carnivore and herbivore groups are
well-defined, including several individual species or pairs of species with
distinct diets. Plant groups demonstrate mild overlap, indicating a partially
hierarchical relationship between smaller groups and larger groups.

Figures~\ref{fig:adjacency_matrix},~\ref{fig:network_aggregated},~and~\ref{fig:network_grouped}
show three alternate views of the food web, organized by the $\kconsensus$-group
consensus partition. Groups $\gTHETRI$--$\gMICKUN$, $\gANDSCH$--$\gACASEN$, and
$\gCRODIC$ consist of plants located in grassland, woodland, and kopje habitat,
respectively. Group $\gTHETRI$ includes short grasses that are grazed by a wide
variety of species, distinguished from group $\gERACIL$, which includes short
grasses fed on only by the large grazers ($\gCONTAU$). The 1-species groups
(\gCANAUR, golden jackal \textit{Canis aureus}; \gSYNCAF, African buffalo
\textit{Syncerus caffer}; \gAGAPLA, agama lizard \textit{Agama planiceps};
\gLOXAFR, elephant \textit{Loxodenta africana}; and \gMICKUN, \textit{Microchloa
kunthii}) represent individual species with distinct diets or consumers.

Plant groups are coupled by groups of herbivores, which are in turn coupled by
groups of carnivores. Large migratory grazers ($\gCONTAU$, wildebeest, zebra,
and gazelles) feed primarily on plant groups consisting of grasses ($\gTHETRI$
and $\gERACIL$), and are the exclusive consumers of a group of grasses adapted
to seasonal rainfall on sandy or volcanic soils ($\gERACIL$) that dominate the
short-grass plains in the southern part of the ecosystem. Herbivores feeding in
the longer grasslands and woodlands and in riparian habitats (group $\gREDRED$)
couple the first grass group ($\gTHETRI$) with woodland plants (group
$\gANDSCH$), which are also consumed to a lesser extent by the large grazers.
The hyraxes (group $\gHETBRU$) and group $\gGIRCAM$ (giraffe, olive baboon, and
dik-dik) couple kopje habitat (group $\gCRODIC$) with both woodland and
grassland plants. At the highest trophic level, the large carnivores
($\gCROCRO$) integrate across all the herbivore groups; smaller carnivores
($\gCANAUR$, $\gCARCAR$) show more specialized diets, reflecting the more
distinct habitats in which they are usually found.

\subsection*{Discussion}

\subsubsection*{Spatial Guilds in the Serengeti Food Web}

In order to analyze the group structure of the Serengeti food web, we used a
flexible Bayesian model of network structure that includes no biological
information aside from a set of nodes representing species and links
representing their interactions. The groups that emerge from an otherwise blind
classification of species make remarkable biological sense. Species are divided
into trophic guilds that reveal a clear relationship between the spatial
organization of plant, herbivore, and carnivore groups and the structure of the
network. At the coarsest scale, the groups in the Serengeti food web correspond
to carnivores, herbivores, and plants. The further subdivisions that emerge
within carnivores, herbivores, and plants reveal a spatial dimension to feeding
structure that is only evident because of high species resolution at the plant
level. Ultimately, the group structure we have derived mirrors the flow of
energy up the food web from different spatial locations, with herbivores
integrating spatially separated groups of plants, and carnivores integrating
spatially widespread herbivores. Although the addition of birds, reptiles,
invertebrates, and pathogens will certainly add a significant number of new
groups, we do not expect them to significantly modify the derived structure for
the mammal and plant community. Nor will they modify the larger tendency for
groups to be assembled in ways that reflect the underlying spatial and trophic
structure of the species in the web.

Recently, interesting theoretical and empirical work has highlighted the
relationship between observed patterns of food-web structure and energy flow
that seemingly mirrors the trophic guild structure in the Serengeti. Rooney and
colleagues \cite{rooney2008} give evidence that real ecosystems may be dominated
by nested sets of fast and slow ``energy channels,'' each of which represents a
food chain of trophic guilds. They suggest that this pattern may have a strong
stabilizing effect, based on theoretical work by McCann on spatially coupled
food webs \cite{mccann2005}. The group structure for the Serengeti web that
emerges from our analysis supports a pattern of spatial coupling at multiple
trophic levels: the grasslands have very high turnover rates compared to those
of the kopjes and woodlands. This suggests a similar pattern of fast and slow
energy channels to those described by Rooney and colleagues, with fast energy
flow up through the highly seasonal but very productive grasses of the
short-grass plains. These are almost completely consumed by wildebeest and zebra
during their peak calving season, which are then in turn consumed by large
predators (lions and hyenas). In contrast, the resident herbivore species living
on kopjes or in the woodlands reproduce at slower rates and are consumed less
frequently by large carnivores, except during the time when they are unable to
feed on migratory wildebeest and zebra. The group model and the inference
approach presented here allow the examination of the dynamical consequences of
this type of structure to be fully rooted in an empirical pattern, complementing
more theoretical considerations of its central importance for preserving
biodiversity \cite{mccann2005}.

These patterns emerge directly from the topology of the food web without being
explicitly labeled as different habitats upfront as was done in previous
empirical work \cite{rooney2008}, showing that topological analysis can reveal
structures that may be very significant for food-web dynamics. They are subtly
different, however, from the proposed pure fast and slow chains, in that they
incorporate the migration of the keystone species in the ecosystem, so the
fastest energy chain is seasonally ephemeral and may only operate for three to
four months in any year. We suspect that even within the sub-habitats of kopjes
and woodlands there are similarly nested faster and slower chains that involve
species for which we are still collating data (e.g., birds, small mammals, and
insects).

\subsubsection*{Bayesian Analysis of Food-Web Structure}

In this paper, we used a probabilistic model to analyze the structure of a
single food web, an approach we have seen in only one other study based on a
probabilistic version of the niche model \cite{williams2010}. This approach has
proved fruitful in Bayesian phylogenetics, where the combinatorial challenges
are similar. Moreover, we view the group model as only a starting point for
richer modeling efforts to help identify relevant processes that influence the
structure of ecological communities.

In fact, the Bayesian approach described here provides a powerful general
framework for encoding hypotheses about the structure of food webs and comparing
models against each other, and we see it as a natural next step in the current
trend of representing food-web models in a common way. Simple abstract models
such as the niche model and the group model used here act as proxies for the
high-dimensional trait space that determines feeding relationships in an
ecosystem. The identification of actual traits that correspond to groups (or
niche dimensions) is another valuable direction, so far followed primarily by
finding correlations between compartments/groups \cite{rezende2009} or niche
values \cite{williams2000} and traits such as body size or phylogenetic
relatedness. Another approach is to directly incorporate these traits into the
probabilistic models, either as model predictors or as informed priors. Both
kinds of analyses are valuable, but the second kind becomes more approachable in
a general Bayesian modeling framework.

The use of flexible hierarchical priors for model parameters is one
straightforward innovation possible in the Bayesian framework. The number of
groups identified by the model increases dramatically with the use of a flexible
beta prior distribution for link probability parameters. In that model variant,
we effectively introduce two degrees of freedom to the model (the beta
distribution parameters) but dramatically reduce the effective degrees of
freedom of the link probability parameters. Note that we properly penalize
parameters by using the marginal likelihood for model selection, so that the
model selection rep- resents a balance between goodness of fit and model
complexity. Moreover, this structure makes intuitive sense: since most link
probability parameters are simply zero, they should not be penalized. An
alternate approach is to remove and add parameters to the model, but this
hierarchical technique is much easier to implement in practice.

Advanced Markov-chain Monte Carlo methods make it possible to accurately
estimate marginal likelihoods for probabilistic network models. Unlike
information criteria such as AIC, BIC, or DIC, an accurate estimate of the
marginal likelihood provides a direct measurement of goodness of fit that takes
into account the degrees of freedom in a model without making any asymptotic
assumptions about parameter distributions \cite{bolker2008}, and can handle
discrete parameters such as partitioning into groups that are not properly
handled by AIC and BIC.

The Bayesian approach also serves as a means to avoid fundamental issues
inherent in network models with a large parameter space. In a recent study, Good
and colleagues \cite{good2010} examined the properties of module-finding in
networks using the pervasive modularity-maximization approach \cite{newman2006},
finding that even in relatively small networks a large number of good solutions
exist. A maximization algorithm is thus guaranteed to find a single local
maximum of many---possibly even the best one, but certainly not one that
captures the full range of good solutions. This problem arises whether the
quantity to be maximized is a heuristic such as modularity or a likelihood
value. The group model and other parameter-rich models presumably suffer from
similar degeneracy problems. In the present case, we find that every partition
sampled from the posterior distribution for the best-fitting group model variant
is unique. Although MCMC sampling cannot reproduce the full posterior
distribution, it is an important step in the right direction. Philosophical
arguments aside, one of the main reasons for maximizing likelihood or modularity
is simply that a single solution is far more tractable than a distribution. The
consensus partitioning heuristic used here is one attempt at recovering a
simpler object of study (see Methods); more sophisticated approaches will be
welcomed.

The group model, based on the simple notion that groups of species may have
similar feeding relationships to other groups, reveals that trophic guilds are
the topologically dominant type of group in the Serengeti food web. The model
also reveals an interesting relationship between spatial structure and network
structure that corroborates recent ideas on spatial coupling in food webs. A
theoretical study with a dynamical model suggests that this type of structure
may contribute to `stability' in the sense of the persistence of species
\cite{mccann2005}. We are now in a position to examine different aspects of
stability, including robustness to secondary extinctions, based on structures
directly inferred from empirical networks. Although the Bayesian modeling
approach is not new to network analysis in general \cite{hoff2002, park2010}, it
remains relatively rare. The Bayesian group model, and, more importantly, the
general framework for modeling and model selection, naturally extend to other
kinds of biological networks, such as metabolic and regulatory networks
\cite{jeong2000} and networks describing other ecological interactions such as
pollination \cite{bascompte2003}. We advocate this framework as a way to build
stronger ties between hypothesis formulation, model building, and data analysis.


\ifthenelse{\plos}
{}
{\end{multicols}}

\section*{Methods}

\subsection*{Group Model}

We use as a starting point the group model of Allesina and Pascual
\cite{allesina2009}, in which a network of $N$ nodes is partitioned into $K$
groups. The groups to which a potential prey and to which a potential predator
belong completely determine the probability that a feeding relationship exists
between them. The assignment of species to groups is given by the vector
$\mathbf{G} = (g_1,\dots,g_n)$, with $g_i \in \{1,\dots,K\}$. We refer to this
assignment as a set `partition,' in keeping with standard mathematical
terminology. The probability that a species assigned to group $i$ is consumed
by a species assigned to group $j$ is equal to $p_{ij}$. This gives a matrix
$\mathbf{P}$ of $K^2$ probabilities, containing the probabilities of observing
directed links between members of each pair of groups, and within members of
each group.

If we take $\mathbf{A}$ to be the directed adjacency matrix of a network, with
entries $a_{ij}$ equal to 1 if a link exists from node $i$ to node $j$, 0
otherwise, then the probability of the network being generated by partition
$\mathbf{G}$ and link probabilities $\mathbf{P}$ is given by
\begin{align}
f(\mathbf{A} | \mathbf{G}, \mathbf{P}) &= \prod_{i=1}^K \prod_{j=1}^K
p_{ij}^{Y_{ij}} (1-p{ij})^{Z_{ij}} \, ,
\end{align}
where $Y_{ij}$ and $Z_{ij}$ are the number of 1-entries and 0-entries in the
submatrix of $\mathbf{A}$ containing entries from rows $r$ satisfying $g_r = i$
and columns $c$ satisfying $g_c = j$.

In the simplest case, all nodes are assigned to the same group, and the likelihood simplifies to
\begin{align}
f(\mathbf{A} | p) &= p^Y (1-p)^Z
\end{align}
where $Y$ and $Z$ are the total number of 1-entries and 0-entries in $\mathbf{A}$.

In order to use the group model for Bayesian inference, we want to
infer the posterior distribution over partitions and parameters,
\begin{align}
f(\mathbf{G}, \mathbf{P} | \mathbf{A}) \propto f(\mathbf{G}, \mathbf{P})
f(\mathbf{A} | \mathbf{G}, \mathbf{P}).
\end{align}
This requires specifying a prior distribution over partitions
$\mathbf{G}$ and link probabilities $\mathbf{P}$. We consider two
priors over $\mathbf{G}$ and two priors over $\mathbf{P}$.

\subsection*{Model Priors}

\subsubsection*{Priors for Partitions}

The simplest prior over partitions assigns equal probability to each
possible assignment of nodes into groups. For a network of $N$ nodes,
the number of possible partitions is given by the $N$th Bell number,
\begin{align}
  \bell(N) &= \sum_{K=1}^N \stirling_2(N, K),
\end{align}
where $\stirling_2(N, K)$ is the Stirling number of the second kind,
the number of ways to partition $N$ objects into exactly $K$ groups,
\begin{align}
  \stirling_2(N, K) &= \frac{1}{K!} \sum_{j=0}^K (-1)^{K-j}
           {\binom{K}{j}} j^N.
\end{align}
Therefore, the prior probability of a particular partition is uniform
across all possible partitions
\begin{align}
  f(\mathbf{G}) &= \frac{1}{\bell(N)},
\end{align}
and the prior probability of having exactly $K$ groups is
\begin{align}
  f(K) &= \frac{\stirling_2(N, K)}{\bell(N)}.
\end{align}
For partitions, the choice of a uniform prior, although simple,
includes hidden assumptions. In particular, there are far more
possible partitions for an intermediate number of groups than a small
or large number, so the prior will implicitly bias results toward that
number. For example, with 100 nodes, the distribution is peaked at $K=28$
(Figure~\ref{fig:k_uniform_prior}).

An alternate prior for partitioning objects into groups comes from the Dirichlet
process, also known as the ``Chinese restaurant process,'' which is becoming a
standard Bayesian prior for related problems
\cite{ferguson1973,huelsenbeck2007,xing2007}. Consider a restaurant with an
infinitely large number of infinitely large tables, all initially empty. The
first patron sits alone at the first table, and subsequent patrons may either
sit at an occupied table or a new table. They choose occupied tables with weight
equal to the number of current occupants, or a new table with weight equal to an
aggregation parameter $\chi$. For example, the second patron will sit at the
same table as the first patron with probability $1/(1 + \chi)$. In fact,
because the process is \textit{exchangeable}, the probability of any pair of
patrons sitting at the table is also $1/(1 + \chi)$. If $\chi$ is small, there
will tend to be a small number of occupied tables and a skewed distribution of
table sizes; if $\chi$ is large, there will be a larger number of tables
occupied by few patrons.

Interpreting tables of patrons as groups of nodes, under the Dirichlet
process the prior probability of a particular partition $\mathbf{G}$
is
\begin{align}
  f(\mathbf{G}|\chi) &= \chi^K \frac{\prod_{j=1}^K (\eta_j -
    1)!}{\prod_{i=1}^N (\chi + i - 1)},
\end{align}
where $N$ is the number of nodes in the network, $K$ is the number of
groups in the partition, and $\eta_j$ is the number of nodes in group
$j$. The prior probability of $K$ groups is
\begin{align}
  f(K|\chi) &= \frac{|\stirling_1(N, K)| \chi^K}{\prod_{i=1}^N (\chi +
    i - 1)},
\end{align}
where $\stirling_1(N, K)$ is a Stirling number of the first kind,
equal to the coefficients on $x_K$ in the expansion $x(x-1)(x-2)
\ldots (x-K+1)$.

Rather than choosing a fixed value of $\chi$ for the prior, we give
$\chi$ an exponential hyperprior distribution with mean $1$:
\begin{align}
  f(\chi) &= e^{-\chi} \qquad \chi \geq 0.
\end{align}

\subsubsection*{Priors for Link Probabilities}

Similarly, the elements of link probability matrix $\mathbf{P}$ may be
given a simple uniform prior over $[0,1]$:
\begin{align}
  f(p_{ij}) &= 1 \qquad 0 \leq p_{ij} \leq 1.
\end{align}

As there may be some regularity in the values of the link
probabilities, we also tried a beta prior:
\begin{align}
  f(p_{ij}|\alpha, \beta) &= \frac{1}{B(\alpha, \beta)} p_{ij}^{\alpha
    - 1} (1 - p_{ij})^{\beta - 1},
\end{align}
where $B(\alpha, \beta)$ is the beta function,
\begin{align}
  B(\alpha, \beta) &= \int_0^1 t^{\alpha - 1} (1-t)^{\beta - 1} \, dt.
\end{align}
The parameters $\alpha$ and $\beta$ control the shape of the
distribution, which may be convex, concave, or skewed toward $0$ or
$1$. When $\alpha = \beta = 1$, the beta prior becomes a uniform
distribution.

We use $\alpha$ and $\beta$ exponential hyperpriors with mean $1$:
\begin{align}
  f(\alpha) &= e^{-\alpha} \qquad \alpha \geq 0, \\ 
  f(\beta) &=  e^{-\beta} \qquad \beta \geq 0.
\end{align}

\subsection*{Markov-chain Monte Carlo Sampling}

For networks of any appreciable size, the number of possible
partitions is far too large to enumerate, so we must use a
Markov-chain Monte Carlo (MCMC) technique to sample from the posterior
distribution. Here we describe the sampling procedure and the details
of the Metropolis-Hastings proposal distribution used.

\subsubsection*{Sampling Procedure}

We employ the standard Metropolis-Hastings algorithm to sample from
the posterior distribution over partitions and model hyperparameters
\cite{metropolis1953,hastings1970}. The general idea of an MCMC method
is to set up a sequence of dependent samples $\theta_1, \theta_2,
\ldots$ that is guaranteed to converge to a target distribution, in
this case the posterior distribution of our model. Starting from the
current sample, a change is proposed, drawn from a \textit{proposal
  distribution} over possible changes, $q(\theta \rightarrow
\theta^*)$. This sample is either rejected, in which case the current
sample is repeated, or the proposed sample is accepted as the new
sample. The Metropolis-Hastings acceptance probability,
\begin{align*}
r(\theta \rightarrow \theta^*) &= \min\left\{1,
	\frac{f(\theta^*)}{f(\theta)}
	\frac{q(\theta^* \rightarrow \theta)}{q(\theta \rightarrow \theta^*)}
	\right\},
\end{align*}
guarantees that the sequence of samples will converge to the posterior
distribution, $f(\theta|D) \propto f(\theta) f(D|\theta)$, the prior
times the likelihood.

For the group model, the samples $\theta$ consist of hyperparameters
for the model variant---the Dirichlet process prior parameter $\chi$
and the beta prior parameters $\alpha$ and $\beta$---as well as the
group count $K$ and assignment vector $\mathbf{G}$. The link
probabilities $\mathbf{P}$ governing links between groups are not
included, because the likelihood function would not be compatible
between partitions with different values of $K$. One possible solution
to this problem would be to include $\mathbf{P}$ in the sampling
procedure, restrict $K$ to a particular number for a particular run,
and then appropriately weight runs with different values of
$K$. Another approach is reversible-jump MCMC \cite{green1995}, which
appropriately handles a mapping between two different parameter spaces
as part of the Metropolis-Hastings proposal ratio. (We tried a reversible-jump scheme, but chains tended to get stuck at local maxima.)

Instead of trying to sample values of $\mathbf{P}$, we use the
\textit{marginal likelihood} of a partition given model
hyperparameters---that is, the posterior distribution, conditional on
values of $\alpha$, $\beta$, and $\mathbf{G}$, integrated over all
possible values of $\mathbf{P}$---directly in the Metropolis-Hastings
procedure. This is possible because the marginal likelihood of a
single partition can be calculated analytically.

For a beta prior over link probabilities, the likelihood of
$\mathbf{G}$, $\alpha$ and $\beta$ marginalized over all possible
values of $\mathbf{P}$ is
\begin{align}
  f(\mathbf{A} | \mathbf{G},\alpha, \beta)
  &= \int_{\mathbf{P}} f(\mathbf{P|\alpha, \beta})
  f(\mathbf{A} | \mathbf{G}, \mathbf{P})
  \, d\mathbf{P} \\
  &= \prod_{i=1}^K \prod_{j=1}^K \int_0^1 \frac{1}{B(\alpha, \beta)}
  p_{ij}^{\alpha - 1} (1 - p_{ij})^{\beta - 1}
  p_{ij}^{Y_{ij}} (1 - p_{ij})^{Z_{ij}} \, dp_{ij} \\
  &= \prod_{i=1}^K \prod_{j=1}^K \int_0^1 \frac{1}{B(\alpha, \beta)}
  p_{ij}^{Y_{ij} + \alpha - 1} (1 - p_{ij})^{Z_{ij} + \beta - 1} \, dp_{ij} \\
  &= \prod_{i=1}^K \prod_{j=1}^K \frac{B(Y_{ij} + \alpha, Z_{ij} + \beta)}{B(\alpha, \beta)}.
\end{align}
Similarly, for a uniform prior over link probabilities, the marginal
likelihood of a particular partition is simply
\begin{align}
  f(\mathbf{A} | \mathbf{G}) &= \prod_{i=1}^K \prod_{j=1}^K B(Y_{ij} + 1, Z_{ij} + 1).
\end{align}

\subsubsection*{Proposal Distribution}

For the case of uniform priors on partitions and link probabilities,
the proposal distribution only allows changes to the partition.  With
the Dirichlet process prior on partitions and the beta prior on link
probabilities, hyperparameters $\chi$, $\alpha$, and $\beta$ can also
be changed.

The proposal distribution is described as follows:
\begin{enumerate}
\item Each hyperparameter $h$ ($\alpha$, $\beta$, and $\chi$) is chosen for
update with probability $p_h$, where $p_h$ are tuned to improve convergence. A
proposed new value $h'$ is drawn from a uniform distribution between $\max(0, h
- r_h)$ and $h + r_h$, where $r_h$ is a proposal radius manually tuned to
improve convergence. (A scale-free proposal could easily be used instead, and
may require less tuning.)
\item With probability $\left(1 - \sum_h p_h \right)$, a group-change move is
proposed:
  \begin{enumerate}
	\item A node $i$ is chosen uniformly at random as the species to be moved.
  	\item Another node $j \neq i$ is chosen uniformly at random.
  	\item If $i$ and $j$ are in different groups, node $i$ is moved into
    the group of node $j$. If $i$ and $j$ are in the same group, node $i$ is
    moved into a new group.
  \end{enumerate}
\end{enumerate}

\subsection*{Metropolis-coupled MCMC}

Although the Metropolis-Hastings algorithm is guaranteed to converge
to the target distribution at some point, local maxima in the
likelihood surface can cause a chain to become stuck for long periods
of time. One approach to avoiding this problem, known as ``Metropolis
coupling,'' involves running multiple chains in parallel.  One chain,
the ``cold chain,'' explores the target distribution, while the other
chains, ``hot chains,'' explore low-likelihood configurations more
freely. Periodically, swaps are proposed between chains, allowing good
configurations discovered on hot chains to propagate toward the cold
chain.

Rather than exploring the target distribution $f(\theta|D) \propto
f(\theta) f(D|\theta)$, heated chains explore
\begin{align}
  f_\tau(\theta|D) \propto f(\theta) \left[ f(D|\theta) \right]^\tau
  \qquad \tau \in [0, 1],
\end{align}
where $\tau$ is a heating parameter. We use linearly spaced values of
$\tau$, with the hottest chain exploring the prior ($\tau = 0$) and
the coldest chain exploring the posterior ($\tau = 1$).

Swap moves are standard Metropolis-Hastings proposals, but rather than
considering a change to a single chain, they consider a change to the
joint distribution of two chains. The acceptance probability is thus
the ratio of the joint distribution after and before the move:
\begin{align}
  r\left((\theta_i, \theta_j) \rightarrow (\theta_j, \theta_i)\right)
  &= \frac{f(\theta_j) \left[f(D|\theta_j)\right]^{\tau_i}
    f(\theta_i) \left[f(D|\theta_i)\right]^{\tau_j}}
  {f(\theta_i) \left[f(D|\theta_i)\right]^{\tau_i}
    f(\theta_j) \left[f(D|\theta_j)\right]^{\tau_j}} \\
  &= \left[\frac{f(D|\theta_i)}{f(D|\theta_j)}\right]^{\tau_j - \tau_i},
\end{align}
where $\theta_i, \theta_j$ are the configurations that begin in chains
$i$ and $j$, and $\tau_i, \tau_j$ are the heat parameters of the two
chains.

The use of multiple heated chains has the side effect of drastically
improving estimates of marginal likelihoods for model selection, as
described in the next section.

\subsection*{Model Selection via Marginal Likelihood}

The marginal likelihood of a model is the likelihood averaged over the
prior distribution. That is, it is the likelihood one would expect by randomly
sampling parameters from the prior distribution:
\begin{align}
  f(D|M) &= \int_\theta f(\theta) f(D|\theta) \, d\theta \, .
\end{align}
This value serves as a useful measure of model fit because it directly
incorporates the dependence of the likelihood on uncertainty in
parameter values, implicitly penalizing extra degrees of freedom
\cite{bolker2008}. If an additional parameter improves the
maximum likelihood but decreases the average likelihood, the model
suffers from overfitting relative to the simpler model.

\subsubsection*{Marginal Likelihood for the Group Model}

The marginal likelihood for the group model involves integrating over
all hyperparameters, partitions, and link probabilities. For the model
with uniform distributions over partitions and link probabilities, the
marginal likelihood is
\begin{align}
  f(\mathbf{A}|M_{u,u}) &= \sum_\mathbf{G} f(\mathbf{G})
  f(\mathbf{A}|\mathbf{G}) \\ &= \sum_\mathbf{G}
  \frac{1}{\bell(N)}
  \left[
    \prod_{i=1}^K \prod_{j=1}^K
    B(Y_{ij} + 1, Z_{ij} + 1)
    \right].
\end{align}
With a Dirichlet process prior over partitions and a uniform distribution over
link probabilities, the marginal likelihood is similarly
\begin{align}
  f(\mathbf{A}|M_{d,u}) &= \int_0^\infty f(\chi) \sum_\mathbf{G}
  f(\mathbf{G}|\chi) f(\mathbf{A}|\mathbf{G}) \, d\chi.
\end{align}
Using a uniform prior over partitions and a beta prior over link
probabilities yields
\begin{align}
  f(\mathbf{A}|M_{u,b}) &= \sum_\mathbf{G} f(\mathbf{G}) \int_0^\infty
  f(\alpha) \int_0^\infty f(\beta) f(\mathbf{A}|\mathbf{G},\alpha,\beta) \,
  d\beta \, d\alpha.
\end{align}
Combining both gives
\begin{align}
  f(\mathbf{A}|M_{d,b}) &= \int_0^\infty f(\chi) \sum_\mathbf{G}
  f(\mathbf{G}|\chi) \int_0^\infty f(\alpha) \int_0^\infty f(\beta)
  f(\mathbf{A}|\mathbf{G},\alpha,\beta) \, d\beta \, d\alpha \, d\chi.
\end{align}

\subsubsection*{Thermodynamic Integration for Marginal Likelihood Estimation}

As enumeration across all possible partitions is impossible for
networks of any significant size, we would like to use MCMC to
estimate the marginal likelihood for the sake of comparison among
different models. Marginal likelihood estimates derived from a single
chain, such as the harmonic mean estimator of Raftery \cite{kass1995},
converge very slowly, because MCMC fails to sample sufficiently from
low-likelihood areas. However, it is possible to use the information
gathered about low-likelihood areas in heated chains using a technique
called thermodynamic integration \cite{lartillot2006,beerli2010}, or
path sampling \cite{gelman1998}.

Assuming a continuum of heated chains, the thermodynamic estimator for
the log-marginal likelihood is
\begin{align}
  \log \hat{\mathcal{L}}(M) &=
  \int_0^1 \frac{1}{m}\sum_{i=1}^m
  \pi(\theta_{i,\tau})
  \log\like(\theta_{i,\tau}) \, d\tau
\end{align}
where $m$ is the number of samples in the MCMC output, and
$\theta_{i,\tau}$ is a single sample from the output in a chain with
heat parameter $\tau$ \cite{beerli2010}. With a finite number of chains, we
estimate this integral using cubic spline interpolation as implemented
in the \texttt{splinefun} function in the R software package \cite{R}.

\subsection*{Consensus Partitions}

The full output of an MCMC chain from the group model includes an
extremely large number of different partitions, and, for the sake of
interpretation, it is desirable to seek a \textit{consensus partition}
that does a reasonable job of summarizing the distribution. We use a
simple, computationally inexpensive method to accomplish this task: in
short, clustering the nodes in the network based on a pairwise
group-membership matrix.

The group-membership matrix $\mathbf{M}$ is the posterior probability
that two nodes are in the same group and 0 otherwise, that is,
\begin{align}
  \mathbf{M} &= \sum_\mathbf{G} P(\mathbf{G} | \mathbf{A})
  \mathbf{M}_\mathbf{G} \, ,
\end{align}
where an entry $\mathbf{M}_\mathbf{G}$ is 1 if nodes $i$ and $j$ are
in the same group, that is,
\begin{align}
  \mathbf{M}_{\mathbf{G},ij} &= \delta_{\mathbf{G}_i, \mathbf{G}_j} \, ,
\end{align}
where $\delta$ is the Kronecker delta and $\mathbf{G}$ is the
assignment vector for the partition. This matrix is estimated from
MCMC output as the fraction of MCMC samples in which the corresponding
species are in the same group:
\begin{align}
  \mathbf{\hat{M}} &= \frac{1}{N}\sum_{i=1}^N
  \mathbf{M}_{\mathbf{G}_i} \, .
\end{align}

A consensus partition is formed by applying a hierarchical clustering
algorithm to the group-\linebreak{}membership matrix estimate $\mathbf{\hat{M}}$,
and then cutting the dendrogram at some number of groups $K$, forming
a consensus partition with assignment vector $\mathbf{G}_K$ and
group-membership matrix $\mathbf{M}_K$. The goodness of fit of a
consensus partition is simply measured as the correlation between
$\mathbf{\hat{M}}$ and $\mathbf{M}_K)$. The best consensus partition
is thus identified using the value of $K$ that gives the highest
correlation.

We use the average-linkage clustering algorithm \cite{sokal1958} as implemented
by the \texttt{hclust} function in the R software package \cite{R}, treating
$\mathbf{1} - \mathbf{\hat{M}}$ as distance matrix. We find that the
average-linkage algorithm produces higher correlations than the other algorithms
implemented as well as ideal $K$ close to the mean $K$ in the MCMC output.
Furthermore, we find that consensus partitions produce higher correlations with
the $\mathbf{\hat{M}}$ than any individual partition in the MCMC output.

\section*{Acknowledgments}

\ifthenelse{\plos}
{A.D. acknowledges the Frankfurt Zoological Society for logistical support in the Serengeti for his work on food webs, and the members of Serengeti Biocomplexity Project for many interesting discussions about the Serengeti food web.}
{We acknowledge the support of NSF grant EF-0827493 (Program on Theory in
Biology) to S.A. and M.P., and of the DOE Computational Science Graduate
Fellowship (grant DE-FG02-97ER25308) to E.B. T.B. was supported by the Howard
Hughes Medical Institute, and M.P. is a Howard Hughes Medical Institute
Investigator.

A.D. acknowledges the McDonnell Foundation for financial support, the Frankfurt
Zoological Society for logistical support in the Serengeti for his work on food
webs, and the members of Serengeti Biocomplexity Project for many interesting
discussions about the Serengeti food web.

The funders had no role in study design, data collection and analysis, decision
to publish, or preparation of the manuscript.}

\bibliography{references}

\ifthenelse{\plos}
{\section*{Figure Legends}}
{\section*{Figures}}

\begin{figure}[!ht]
	\begin{center}
		\ifthenelse{\plos}
		{}
		{\includegraphics[width=4in]{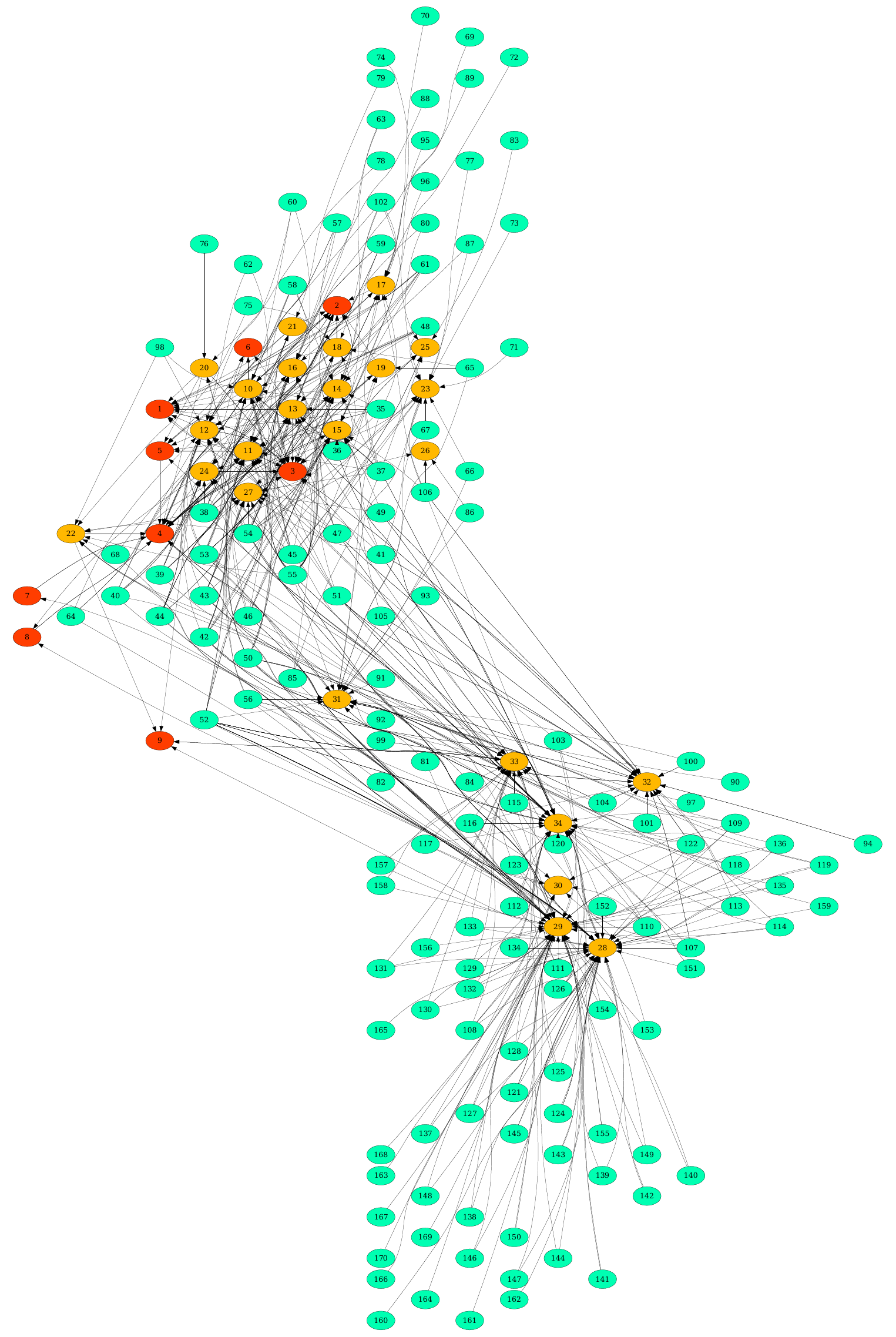}}
	\end{center}
	\caption{{\bf The Serengeti food web.} The network is shown using a
	spring-layout algorithm without clustering. Plant, herbivore, and carnivore
	nodes are green, orange, and red, respectively.  }
	\label{fig:network_ungrouped}
\end{figure}

\begin{figure}[!ht]
	\begin{center}
		\ifthenelse{\plos}
		{}
		{\includegraphics[width=4in]{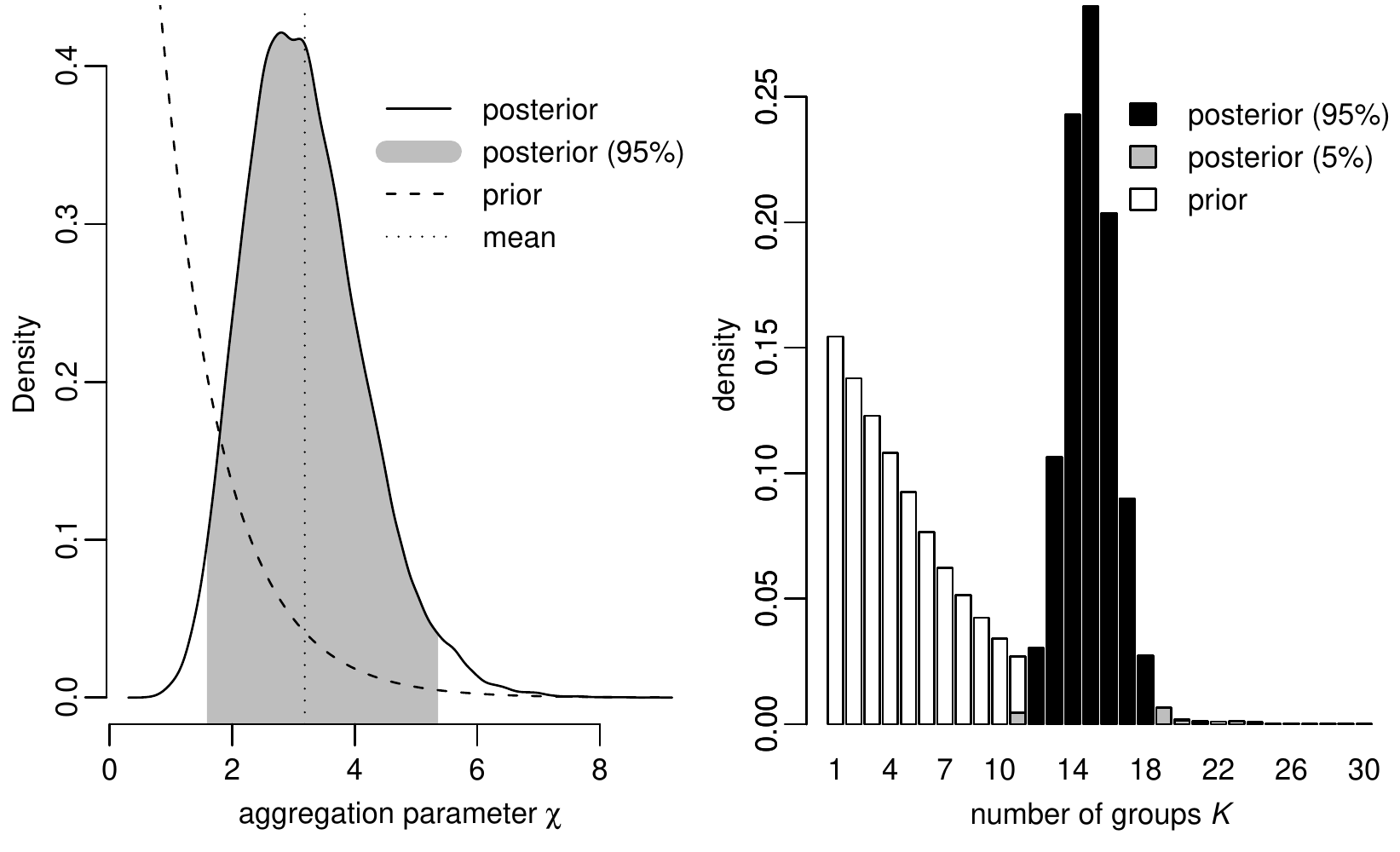}}
	\end{center}
	\caption{{\bf Posterior distributions and prior expectations of aggregation
	parameter $\chi$ and group count $K$.} }
	\label{fig:chi_k}
\end{figure}

\begin{figure}[!ht]
	\begin{center}
		\ifthenelse{\plos}
		{}
		{\includegraphics[width=4in]{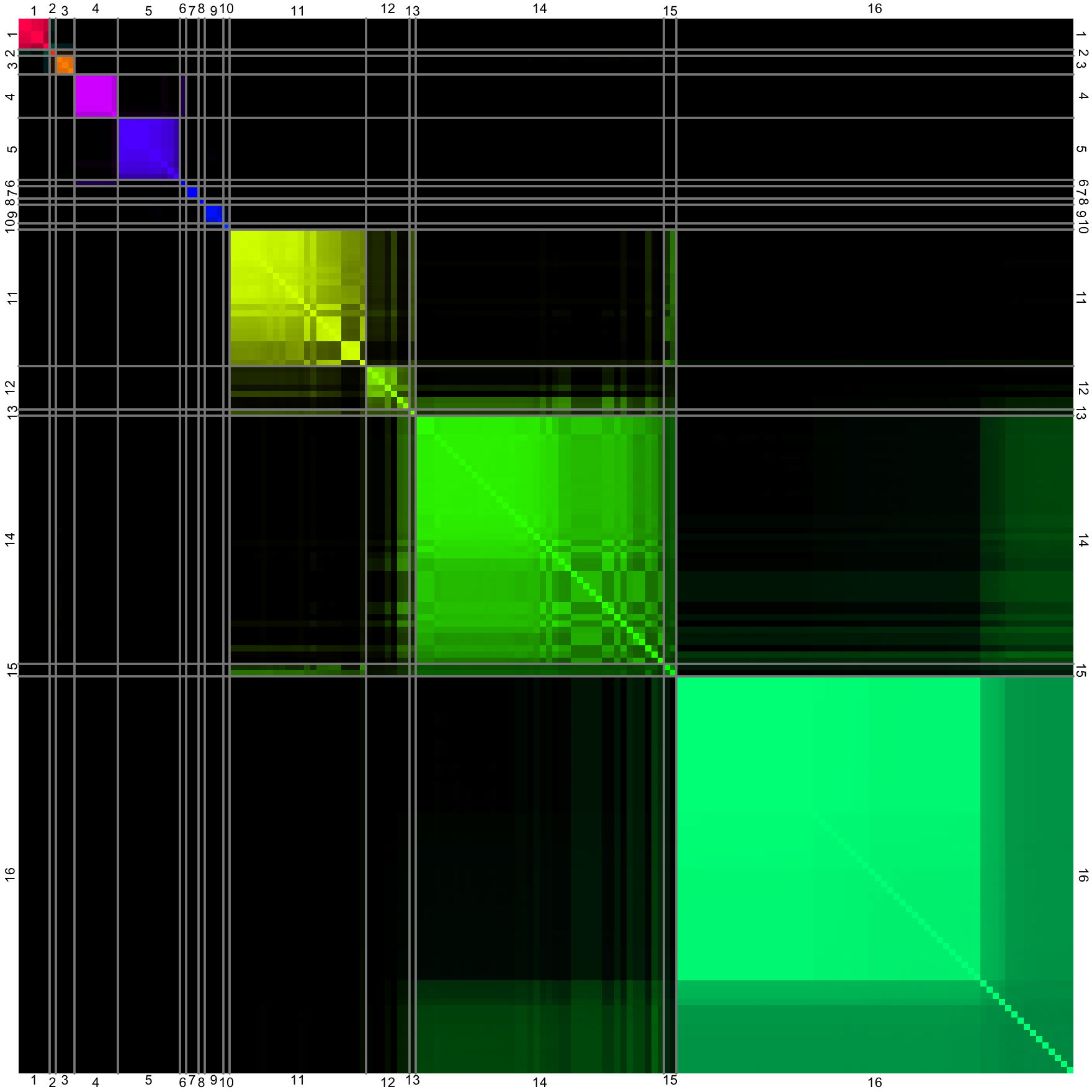}}
	\end{center}
	\caption{
	{\bf Pairwise group membership matrix.} Species are identically
	ordered top to bottom and left to right according to the $\kconsensus$-group
	consensus partition as listed in Table~\ref{tab:groups}. Hue indicates
	group identity; color saturation indicates the fraction of partitions in which
	species occupy the same group. Note that this image conveys information about
	group membership, not network connectivity.}
	\label{fig:similarity_matrix}
\end{figure}

\begin{figure}[!ht]
	\begin{center}
	\ifthenelse{\plos}
	{}
	{\includegraphics[width=1.5in]{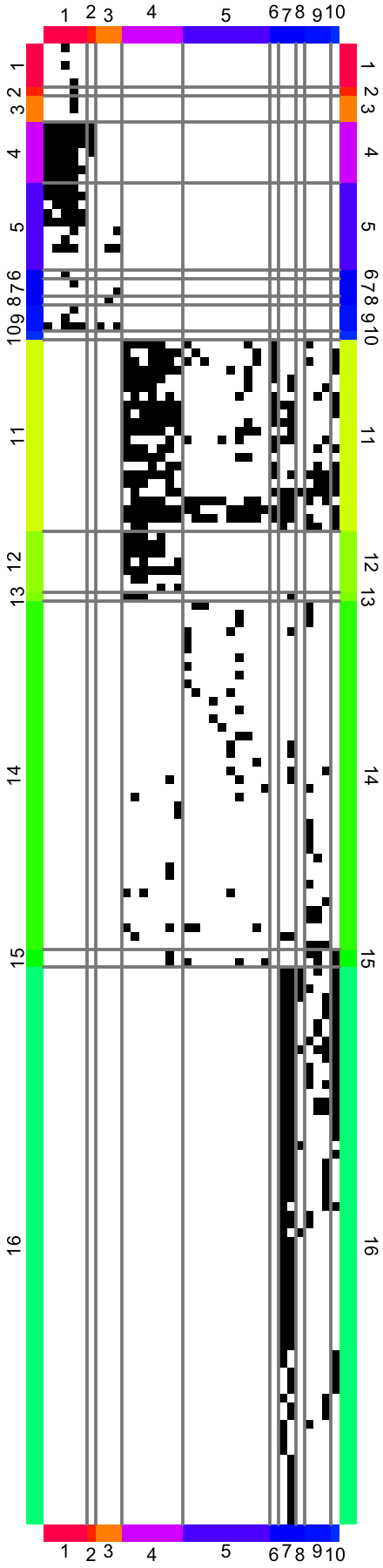}}
	\end{center}
	\caption{
	{\bf Adjacency matrix ordered by groups.} Species are identically ordered top to
	bottom and left to right according to the $\kconsensus$-group consensus
	partition as listed in Table~\ref{tab:groups}. Black matrix entries indicate
	that the species in the column feeds on the species in the row. Columns that
	would indicate prey of plant groups are omitted. Note that in a modular network
	according to the standard definition, links would be concentrated on the
	diagonal of the adjacency matrix, since they occur within groups. By contrast,
	here links are concentrated in off-diagonal blocks.}
	\label{fig:adjacency_matrix}
\end{figure}

\begin{figure}[!ht]
\begin{center}
	\ifthenelse{\plos}
	{}
	{\includegraphics[width=5in]{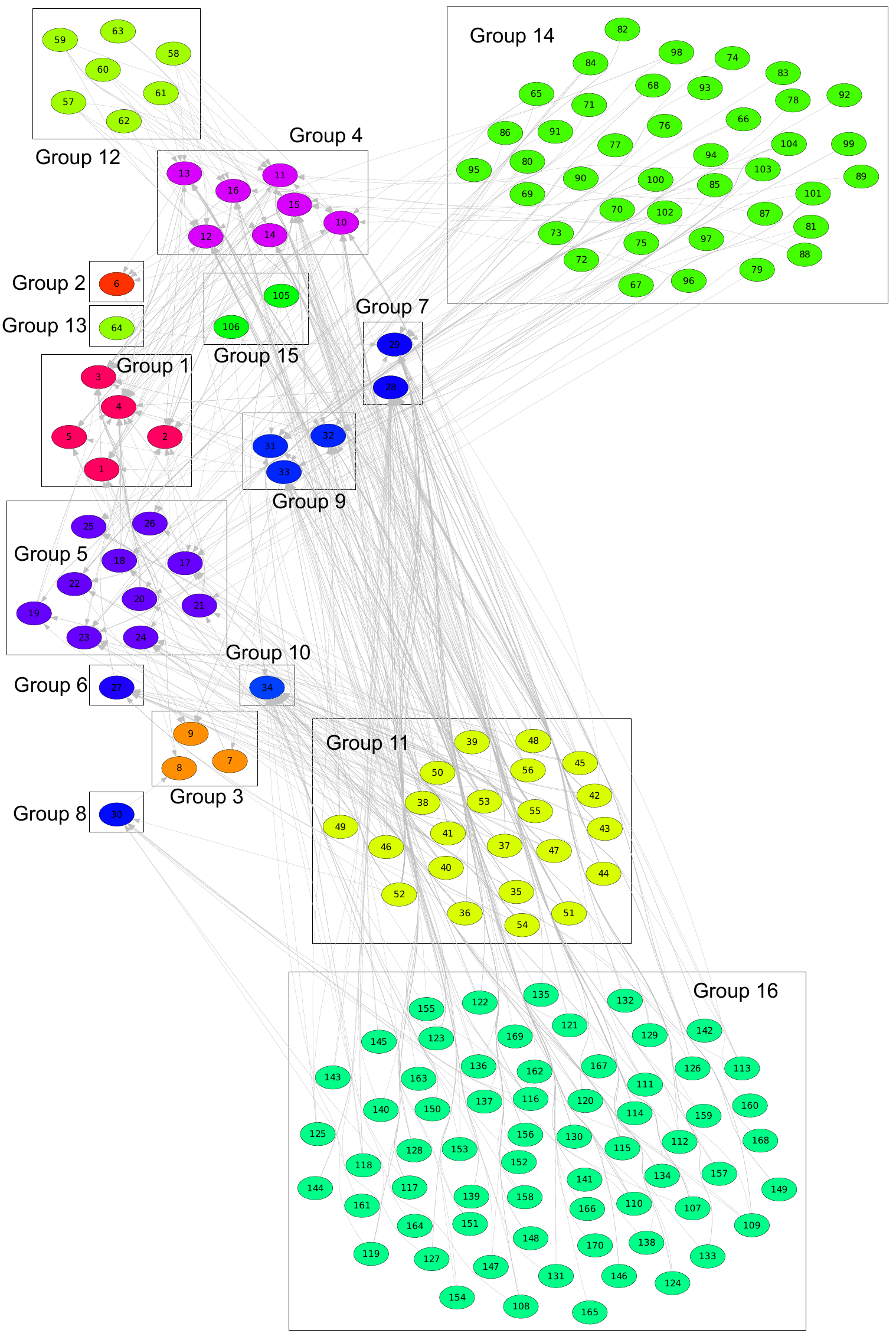}}
\end{center}
\caption{
{\bf Network layout of groups.} The network is shown organized and colored by
group according to the $\kconsensus$-group consensus partition listed in Table~\ref{tab:groups}.}
\label{fig:network_grouped}
\end{figure}

\begin{figure}[!ht]
\begin{center}
	\ifthenelse{\plos}
	{}
	{\includegraphics[width=5in]{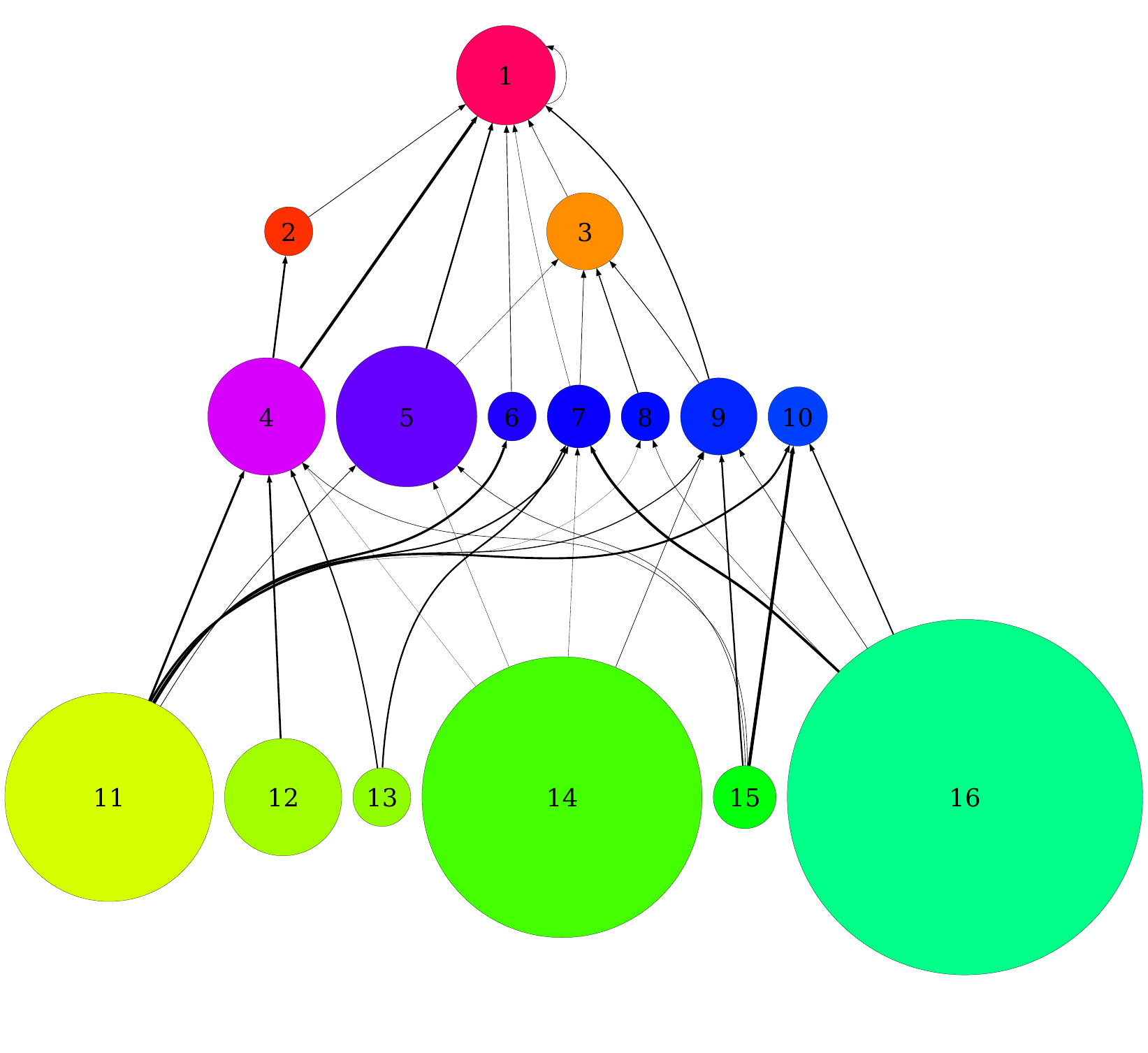}}
\end{center}
\caption{
{\bf Network layout of aggregated groups.} Nodes in the network are aggregated
and colored by group according to the $\kconsensus$-group consensus partition
listed in Table~\ref{tab:groups}, and arranged vertically by trophic level. Line
thickness indicates the link density between groups. Node area is proportional
to the number of species in a group.}
\label{fig:network_aggregated}
\end{figure}


\pagebreak

\section*{Supporting Information}
\setcounter{figure}{0}
\setcounter{table}{0}
\renewcommand{\thefigure}{S\arabic{figure}}
\renewcommand{\thetable}{S\arabic{table}}

\begin{figure}[!ht]
	\begin{center}
		\ifthenelse{\plos}
		{}
		{\includegraphics[width=4in]{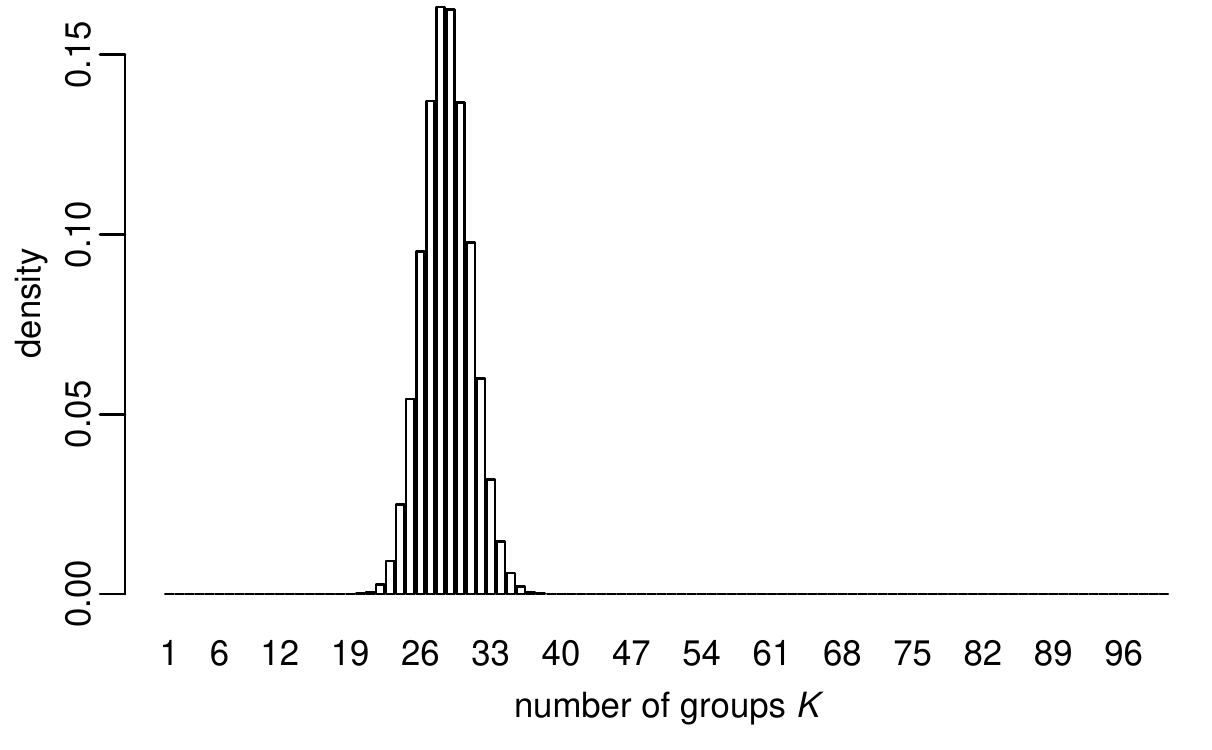}}
	\end{center}
	\caption{{\bf Implicit prior on number of groups with uniform partition prior.}
	The prior distribution on number of groups $K$ is shown for a uniform
	partition prior for a network with 100 nodes. The mode of the distribution is
	at $K = 28$.}
	\label{fig:k_uniform_prior}
\end{figure}

\begin{figure}[!ht]
	\begin{center}
		\ifthenelse{\plos}
		{}
		{\includegraphics[width=4in]{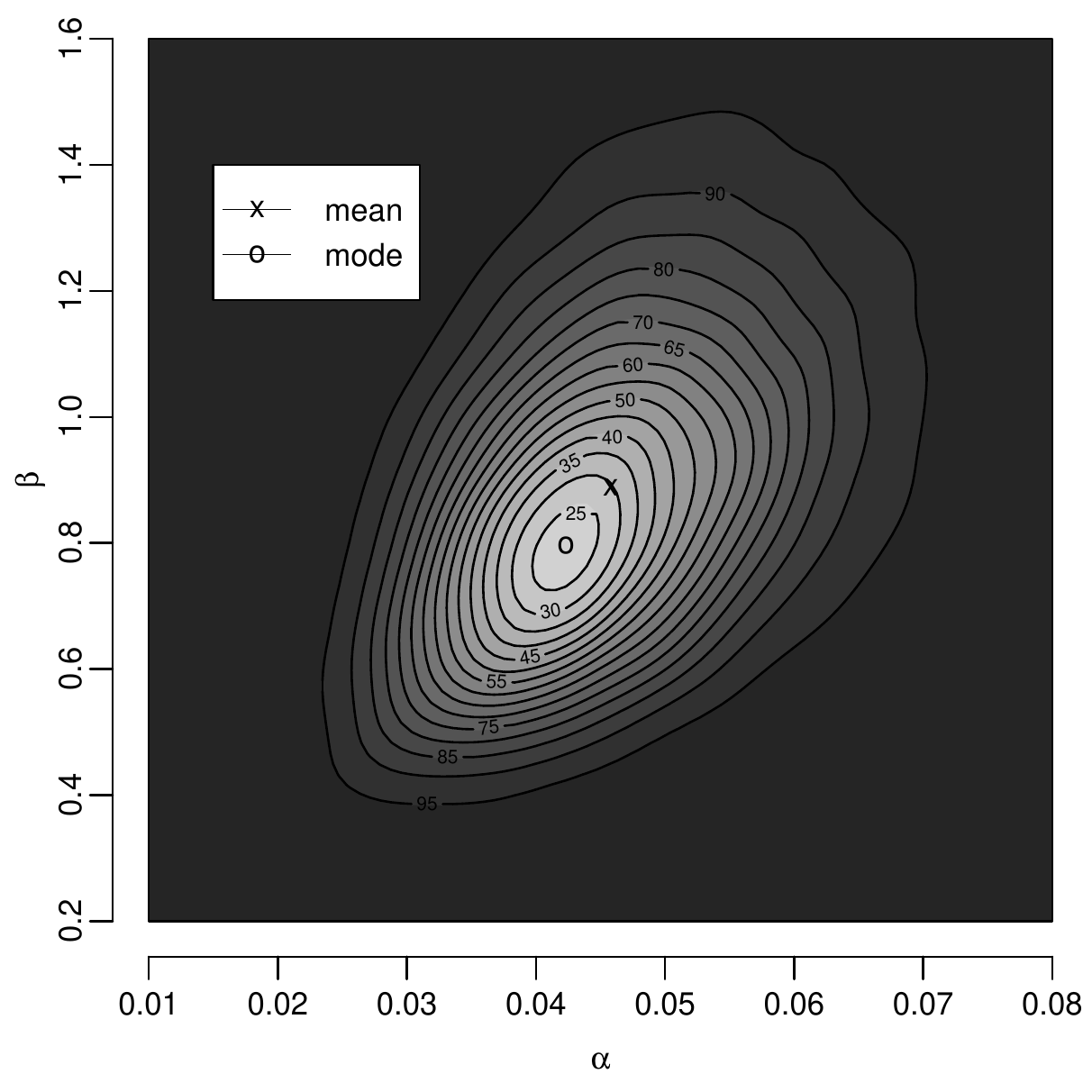}}
	\end{center}
	\caption{{\bf Posterior distributions of link density parameters $\alpha$ and
	$\beta$.} Color brightness indicates posterior density, estimated using the
	\texttt{ks} multivariate kernel density estimation package for R
	\cite{duong2007}. Contours indicate cumulative density. The $\alpha$ parameter is significantly lower than
	1, indicating departure from a uniform distribution.}
	\label{fig:alpha_beta}
\end{figure}

\begin{figure}[!ht]
	\begin{center}
		\ifthenelse{\plos}
		{}
		{\includegraphics[width=4in]{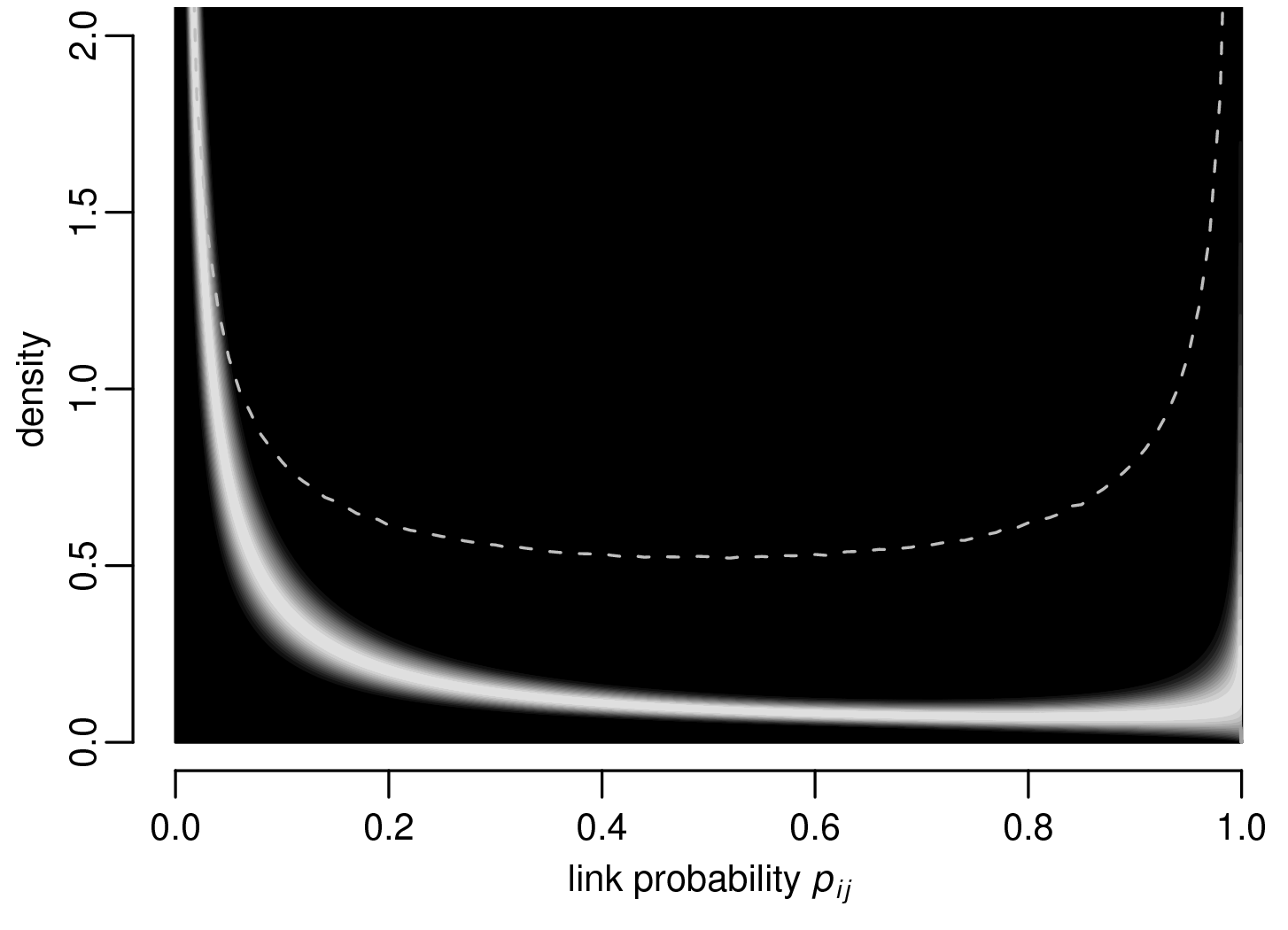}}
	\end{center}
	\caption{{\bf Distribution of link probability parameters.} The prior
	distribution for link probability parameters, integrated over the priors for
	beta distribution parameters $\alpha$ and $\beta$, is indicated with a dotted
	line. The heat map shows beta distributions corresponding to the posterior
	distribution for $\alpha$ and $\beta$, with lightness indicating the posterior
	density of the parameter values.}
	\label{fig:pij}
\end{figure}

\begin{table}[!ht]
	\begin{center}
		\ifthenelse{\plos}
		{}
		{(See supporting file \texttt{tab\_S1\_species\_list.csv} for data.)}
	\end{center}
	\caption{{\bf Species in the Serengeti food web.}}
	\label{tab:species_list}
\end{table}

\begin{table}[!ht]
	\begin{center}
		\ifthenelse{\plos}
		{}
		{(See supporting file \texttt{tab\_S2\_edge\_list.csv}  for data.)}
	\end{center}
	\caption{{\bf Feeding links in the Serengeti food web.}}
	\label{tab:edge_list}
\end{table}

\begin{table}[!ht]
	\begin{center}
		\ifthenelse{\plos}
		{}
		{(See supporting file \texttt{tab\_S3\_consensus\_partition.csv}  for data.)}
	\end{center}
	\caption{{\bf $\mathbf{\kconsensus}$-group consensus partition.}}
	\label{tab:consensus_partition}
\end{table}

\begin{table}[!ht]
	\begin{center}
		\ifthenelse{\plos}
		{}
		{(See supporting file \texttt{tab\_S4\_link\_density.csv} for data.)}
	\end{center}
	\caption{{\bf Link densities between groups in the $\mathbf{\kconsensus}$-group
	consensus partition.}}
	\label{tab:link_density}
\end{table}


\section*{Tables}

\setcounter{figure}{0}
\setcounter{table}{0}
\renewcommand{\thefigure}{\arabic{figure}}
\renewcommand{\thetable}{\arabic{table}}

\begin{table}[!ht]
	\centering
	\caption{
	\bf{Marginal likelihood estimates for model variants calculated via
	thermodynamic integration.}}
	\begin{tabular}{l|l|c|c}
	\hline
	Partition prior & Link prior & Log marginal likelihood estimate & 95\% bootstrap
	confidence int.
	\\
	\hline
	Uniform & Uniform & $-1826.87$ & $(-1826.94, -1826.82)$ \\
Uniform & Beta & $-1463.22$ & $(-1463.39, -1463.03)$ \\
Dirichlet process & Uniform & $-1547.41$ & $(-1547.45, -1547.37)$ \\
Dirichlet process & Beta & $-1356.96$ & $(-1357.02, -1356.90)$ \\
One group & Uniform & $-2870.58$ & $\text{(exact)}$ \\
$\nspecies$ groups & Uniform & $-20031.95$ & $\text{(exact)}$ \\
$\nspecies$ groups & Beta & $-2870.58$ & $\text{(exact)}$ \\
	\hline
	\end{tabular}
	\label{tab:marginal_likelihood}
\end{table}

\begin{table}[!ht]
	\caption{
	\bf{Groups identified in the Serengeti food web using a
	$\mathbf{\kconsensus}$-group consensus partition.}}
	\begin{tabular}{|c|p{5.4in}|c|}
	Group 1 & Crocuta crocuta, Lycaon pictus, Panthera leo, Panthera pardus, Acinonyx jubatus \\
Group 2 & Canis aureus \\
Group 3 & Canis mesomelas, Leptailurus serval, Caracal caracal \\
Group 4 & Connochaetus taurinus, Gazella granti, Gazella thomsoni, Equus burchelli, Alcelaphus buselaphus, Aepyceros melampus, Damaliscus korrigum \\
Group 5 & Kobus ellipsiprymnus, Phacochaerus aethiopicus, Tragelaphus scriptus, Ourebia ourebi, Redunca redunca, Pedetes capensis, Taurotragus oryx, Rhabdomys pumilio, Hippopotamus amphibus, Cercopithecus aethiops \\
Group 6 & Syncerus caffer \\
Group 7 & Heterohyrax brucei, Procavia capensis \\
Group 8 & Agama planiceps \\
Group 9 & Papio anubis, Giraffa camelopardalis, Madoqua kirkii \\
Group 10 & Loxodenta africana \\
Group 11 & Panicum coloratum, Sporobolus pyramidalis, Hyparrhenia filipendula, Harpachne schimperi, Digitaria macroblephara, Eragrostis tenuifolia, Grewia bicolor, Aristida adoensis, Brachiaria semiundulata, Pennisetum mezianum, Bothriochloa insculpta, Panicum maximum, Sida spp., Eustachys paspaloides, Croton macrostachyus, Solanum incanum, Indigofera hochstetteri, Hibiscus spp., Heteropogon contortus, Cynodon dactylon, Themeda triandra, Balanites aegytiaca \\
Group 12 & Digitaria scalarum/abysinnica, Dinebra retroflexa, Ischaemum afrum, Eragostris cilianensis, Hyparrhenia rufa, Sporobolus fimbriatus, Sporobolus spicatus \\
Group 13 & Microchloa kunthii \\
Group 14 & Echinochloa haploclada, Digitaria milanjiana, Panicum deustum, Digitaria ternata, Andropogon schirensis, Cymbopogon excavatus, Setaria sphacelata, Typha capensis, Setaria pallidifusca, Phragmites mauritianus, Eragrostis exasperata, Andropogon greenwayi, Lonchocarpus eriocalyx, Sporobolus centrifugus, Hyparrhenia dissoluta, Chloris roxburghiana, Aristidia hordacea, Chloris pycnothrix, Panicum repens, Aristidia kenyensis, Combretum molle, Acacia xanthophloea, Disperma kilimandscharica, Vossia cuspida, Odyssea jaegeri, Sporobolus ioclados, Euphorbia candelabrum, Sorghum versicolor, Kigelia africana, Olea spp., Sporobolus festivus, Acacia pallens, Crotalaria spinosa, Digitaria diagonalis, Boscia augustifolia, Acacia robusta, Acacia seyal/hockii, Chloris gayana, Pennisetum stramineum, Commiphora africana/trothae \\
Group 15 & Acacia senegal, Acacia tortilis \\
Group 16 & Grewia fallax, Cissus quadrangularis, Cissus rotundifolia, Commelina africana, Allophylus rubifolus, Sensevieria ehrenbergiana, Pavetta assimilis, Phyllanthus sepialis, Acalypha fructicosa, Maerua triphllya, Ficus glumosa, Croton dichogamus, Sclerocarya birrea, Capparis tomentosa, Ximenia caffra, Cordia ovalis, Grewia trichocarpa, Abutilon angulatum, Pappaea capensis, Commiphora schimperi, Albuca spp., Ficus ingens, Hoslundia opposita, Ocinum suave, Cenchrus ciliaris, Solanum dennekense, Aloe macrosiphon, Indigofera basiflora, Ipomoea obscura, Albizia harveyi, Ficus thinningii, Emilia coccinea, Cyphostemma nierensis, Spirocarpa spp., Sensevieria suffruticosa, Pupalia lappacea, Aloe secundiflora, Turreae fischeri, Pavonia patens, Jasminum fluminense, Acacia clavigera, Cassia didymobotrya, Kedrotis foetidissima, Hypoestes forskalii, Zisiphus mucronata, Commiphora merkeri, Blepharis acanthoides, Iboza sp., Rhoicissus revoilii, Kalanchoe sp., Solanum nigrum, Achyranthes aspera, Digitaria velutina, Tricholaena eichingeri, Lippia ukambensis, Heliotropium steudneri, Kyllinga nervosa, Sporobolus stapfianus, Cyperus kilimandscharica, Pellaea calomelanos, Sporobolus pellucidus, Eragrostis aspera, Eriochloa nubica, Diheteropogon amplectus
	\end{tabular}
	\label{tab:groups}
\end{table}

\end{document}